\documentclass[nologo,11pt,a4paper,hidelinks]{ETHpaper}
\usepackage{graphicx, epsfig, amsmath, amssymb,color,wasysym}
\usepackage[numbers]{natbib}
\usepackage{subcaption}
\usepackage{booktabs}
\usepackage{mathtools}
\usepackage{cancel}
\usepackage[table]{xcolor}
\usepackage{colortbl}
\usepackage{algorithm2e}
\usepackage{enumitem}

\definecolor{lgray}{RGB}{210,210,210}

\begin{document}

\newcommand{\mean}[1]{\left\langle #1 \right\rangle} 
\newcommand{\abs}[1]{\left| #1 \right|}

\title{Quantifying the importance of firms \\ by means of reputation and network control
}   

\titlealternative{Quantifying the importance of firms by means of reputation and network control
}   

\author{Yan Zhang, Frank Schweitzer}

\authoralternative{Yan Zhang, Frank Schweitzer}

\address{Chair of Systems Design, ETH Zurich, Weinbergstrasse 58, 8092, Zurich, Switzerland}

\reference{(Submitted for publication)} \www{\url{http://www.sg.ethz.ch}}

\makeframing
\maketitle

\begin{abstract}
  The reputation of firms is largely channeled through their ownership structure.
  We use this relation to determine reputation spillovers between transnational companies and their participated companies in an  ownership network core of 1318 firms.
  We then apply concepts of network controllability to identify minimum sets of driver nodes (MDS) of 314 firms in this network.
  The importance of these driver nodes is classified regarding their control contribution, their operating revenue, and their reputation.
  The latter two are also taken as proxies for the access costs when utilizing firms as driver nodes.
  Using an enrichment analysis, we find that firms with  high reputation maintain the controllability of the network, but rarely become top drivers, whereas firms with medium reputation most likely become top driver nodes.
   We further show that 
  MDSs with lower access costs can be used to control the reputation dynamics in  the whole network.

\end{abstract}

\section{Introduction}
\label{sec:motivation}

Reputation is a precious value for social and economic actors, such as individuals, organizations, or firms. 
To built up reputation may take a long time, but it can be destroyed very quickly.
This asymmetry between growth and decay needs to be taken into account when we wish to model reputation dynamics.
In order to achieve such a model, we first need to think about ways to quantify reputation.
In this paper, we focus on the reputation of \emph{firms}.
Traditionally, corporate reputation is evaluated via surveys. 
This often results in reputation rankings \citep{Fombrun2015}, i.e. a comparison of relative, rather than absolute  reputation.   
This approach makes it quite difficult to compare the reputation of firms at large scale, for instance across different industrial sectors.  
Further, classical reputation rankings do not allow to address the important problem of \emph{reputation spillover}, i.e. the in/decrease of a firm's reputation based on the in/decrease of the reputation of other firms it depends on.
In fact, \emph{reputation risk} has become a major issue in particular for large transnational companies, whose reputation can be drastically hampered by their suppliers, investors, or manufacturing partners \citep{Louisot2004,Brammer2006,Kang2008}.

To overcome the problems of measuring reputation and quantifying reputation spillovers, we have proposed a framework that has been successfully applied to online social networks \citep{Reptuation-OSN-2020}.
The main ideas of our reputation measure are summarized in Section~\ref{sec:dynamics-reputation}.
Here we build on this framework to address a more ambitious question, namely how to \emph{control} the reputation of firms.
This requires us to first clarify what we mean by control.

Nowadays, already the attempt to ``control'' social or economic actors raises ethical or legal concerns.
We do not enter such discussions here.
Instead, we point to two established research directions, network interventions and network controllability, which we also utilize in our paper.
Following these concepts, control means to influence a system such that it obtains a ``better'' state.
In the socio-economic realm, this can be a more resilient state for infrastructure networks, a state with higher capital per capita for countries, or a state of higher trust between individuals.

Systems design distinguishes two approaches to obtain such improvements \citep{Schweitzer:2019vp}.
The \emph{top-down} approach tries to optimize boundary conditions, e.g. tax rates or legal frameworks for \emph{all} firms, to enable a positive development.
The \emph{bottom-up} approach, on the other hand, focuses on system elements, e.g. 
\emph{single} firms, that can be targeted as seeds for a positive development.
In this paper, we are interested in the second approach to improve the state of a system of firms.
That means, we want to influence \emph{individual} firms, to obtain a better \emph{systemic} outcome.

Already classical game theory discusses the option to change either the payoff matrix or the available information such that a particular strategy, e.g. cooperation, becomes more attractive to players.
The concept of \emph{nudging} has build on this, subtly influencing the decision of social or economic actors in favor of a preferred  outcome \citep{Sugden2009}.
Network interventions further leverage this idea by using the interaction network as an amplifier \citep{valente2012network,valente2017putting}.
For example, changing the utility function of a single firm, or a user, impacts other firms and users directly or indirectly via the network \citep{intervention-knowledge-Zhang2020}.
This has proven to be an effective and a cost efficient way to obtain an outcome that is more desirable from the perspective of a social planner \citep{LeoneSciabolazza2020}.
This way, for instance the resilience of social networks could be improved \citep{interventions-OSN-Casiraghi-2020}.

The concept of network interventions requires to know, and to monitor, the system state that should be achieved.
This is very often hard to quantify.
Here, the more abstract concept of \emph{network controllability} comes into play \citep{Liu2011d,Cornelius2013,Wang2012e}. 
It derives from \emph{control theory}, originally developed in engineering and operations research.
Network controllability focuses on the question what part of a network can be controlled if we steer a particular node, or a set of nodes, which are called \emph{driver nodes}.
Control means here that this part of the network can be driven into \emph{any} possible state that is compatible with the assumed network structure and dynamics.
Similar to network interventions,  \emph{not} all nodes in a network shall be targeted, ideally the set of driver nodes is rather small.
But different from network interventions, we do not need to specify the desired system state.
Instead, the principal ability to influence (part of) the network is investigated.

Following this framework in our paper, we can assign each node in the network a ``capacity'' to influence the network.
But not all nodes qualify as driver nodes.
Hence, in a first step we have to identify the set of driver nodes - which becomes a demanding task in particular for large networks, such as the network of firms with 1318 nodes discussed in this paper.
To solve this problem, we need to know (i) the network structure and (ii) the dynamics that couples the nodes, which is the dynamics of \emph{reputation spillover}. 

In Section~\ref{sec:dynamics-reputation} we summarize this dynamics for the \emph{reputation of firms}.
We also introduce the network that we want to leverage for influencing firms, which is their \emph{ownership network}. 
Here we build on a recent study that quantifies the relation between corporate reputation and ownership \citep{Zhang2019}.
Eventually, in Section~\ref{sec:ident-driv-nodes} we summarize the algorithmic procedure to identify the set of driver nodes, 
following the concept of network controllability.

In Section~\ref{sec:results} we present the results of our study.
Our focus is on the question how the control contribution of firms, i.e. their ability to steer the network dynamics, is related to their reputation, as measured by our framework.
Na\"ively, one could assume that the most influential firms, as measured by their control contribution, are the firms with the highest reputation.
This would imply that utilizing such firms as drivers may become a costly endeavor, because of their pronounced economic value.
Our major finding is that this in fact does not hold.
Instead, we could identify a larger number of less reputed firms to drive the network.
This insight can open new ways to influence such economic systems.

\section{Material and Methods}
\label{sec:material-methods}

\subsection{Data set of transnational firms}
\label{sec:data-set-transn}

The availability of large-scale data sets about firms has boosted research about \emph{economic networks} in the recent decade \citep{ACS-econ-netw-2009}. 
To construct such networks, different types of interactions between firms have been analyzed, for example knowledge transfer \citep{reagans2003network,JEE2018}, financial relations \citep{Battiston2012,Nanumyan2015}, supply chains and trade networks \citep{Mizgier2013,fagiolo2010evolution,garlaschelli2005structure,Schweitzer2019}, or ownership \citep{mani2014moving,garcia2017uncovering,vitali-glattfelder-battiston-2011-networ-global,glattfelder-battiston-2009-backb}.

In this paper, we build on the latter, by reusing a data set about the global ownership relations among firms \citep{vitali-glattfelder-battiston-2011-networ-global,glattfelder-battiston-2009-backb,Zhang2019} obtained from the ORBIS data base of 2007. 
This reports information about the share firm $A$ holds on firm $B$, i.e. links in the ownership network are \emph{directed} and \emph{weighed}.
Further information about the operating revenue of each firm is available. 
This data set has been previously analyzed to quantify corporate control \citep{vitali-glattfelder-battiston-2011-networ-global,glattfelder-battiston-2009-backb}. 

Similar to the mentioned works,  in the following we focus on \emph{transnational companies} (TNCs) which, according to the OECD definition, operate in more than one country.
They are known to form the backbone on the ownership network \citep{glattfelder-battiston-2009-backb}. 
These TNCs directly or indirectly participate in other firms, called \emph{participated companies} (PCs) which are mostly direct or indirect subsidiaries of TNCs.

\begin{figure}[htbp]
  \centering
  \includegraphics[width=0.55\textwidth]{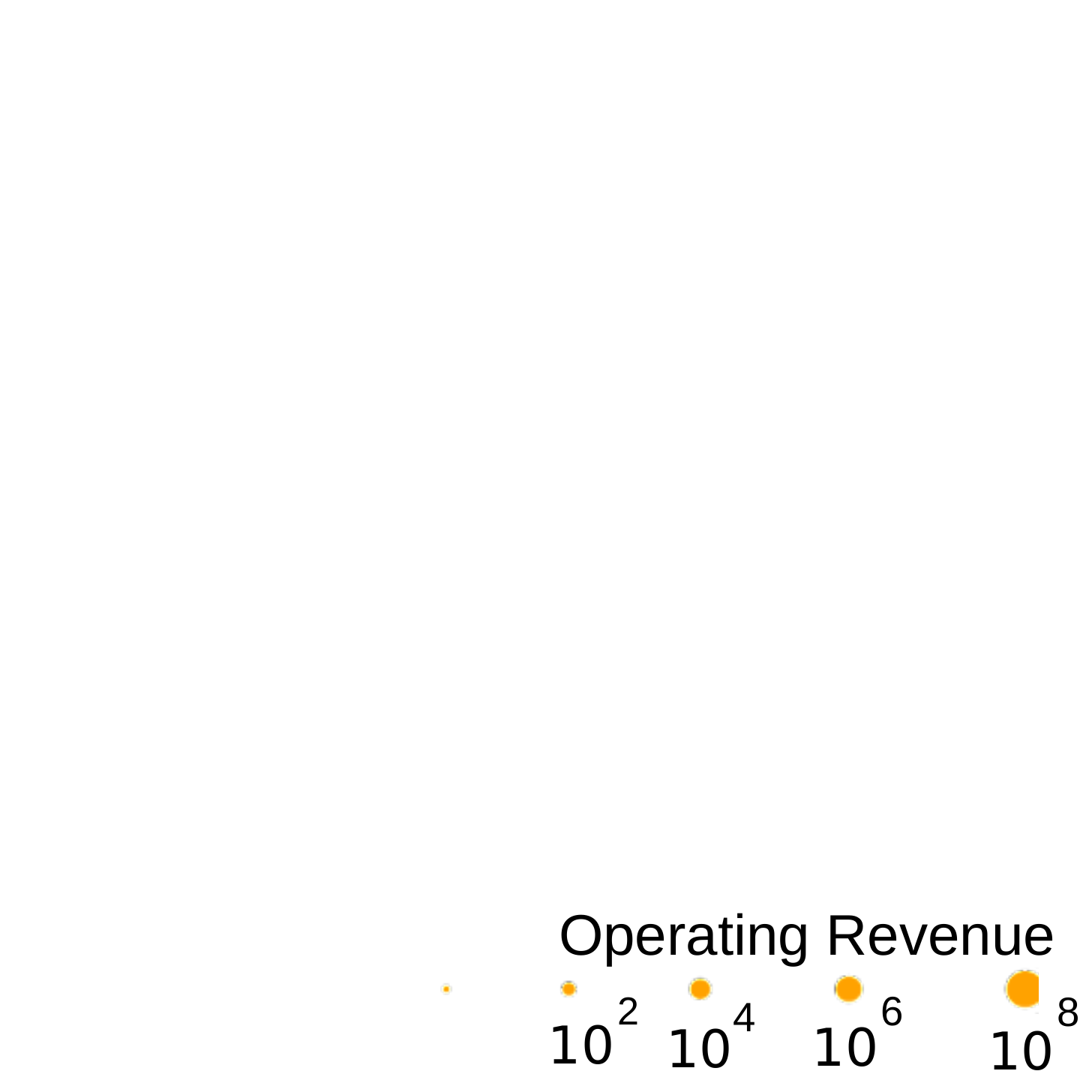}\caption{Visualization of the densely connected core of the global ownership network of 1318 firms.
    Transnational companies (TNCs) are shown in orange, participated companies (PCs) are shown in light green nodes.
    The size of each node is scaled according to the operating revenue of the firm.
      }
\label{fig:ownership}
\end{figure}

Our analyisis is focused on the very small, but densely connected core of this network \citep{vitali-glattfelder-battiston-2011-networ-global} which is also visualized in Figure~\ref{fig:ownership}.
It comprises 1318 firms that are connected by 12184 ownership relations.
I.e.,  on
average, each firm is connected to 20 other firms and there is at least one
directed path from any firm to other firm in this core.
The overall operating revenue for firms in this core accounts for 20\% of the operating revenue by all firms in the global ownership network.
So, we are looking here at the heart of the global economy.

We will use this ownership network to later explore the network controllability by identifying the set of driver nodes.
For this, we also need to specify the dynamics that connect these firms.

\subsection{Dynamics of reputation}
\label{sec:dynamics-reputation}

The ownership relations between firms do not only determine corporate influence, they also channel \emph{reputation} \cite{Fombrun1990,Brammer2006,Delgado-Garcia2010}.
In a recent paper \citep{Zhang2019}, we have distinguished two phases, which differ in the directionality for this reputation spillover.
In an initial phase, the reputation of the owners, i.e. the firms investing into a newly founded company, largely determines the reputation of this firm, because 
with their reputation early shareholders signal trust to invest in this yet unknown firm.

In the second phase, the reputation of the invested firm can feed back on the reputation of its stakeholders, both in positive and negative ways. 
We do not need to list the many scandals that have shaken the business world because reputed stakeholders, which also represent a considerable corporate control, have been made responsible for the malfunction of their firms.
But investors also use the positive reputation of firms, e.g. in the green energy sector, to brush up their own reputation - as the recent debate on ethical investments witnesses.

Thus, it is justified to discuss the reputation dynamics of firms by utilizing their ownership network.
In the following we focus on the core of the ownership network which represents a mature economy of established firms.
Therefore, we consider the second phase, where the directionality  of the ownership links is opposite to the directionality of the reputation spillover.
I.e. reputation spills over from the invested firm to its shareholders. 

To quantify reputation, we assign to each firm a scalar value, $x_{i}(t)$ that changes in time according to the following dynamics \citep{YZ-ACS}:
\begin{align}
  \label{eq:1}
  \dot{x}_{i}(t)=\sum_{j=1}^{N}a_{ji} x_{j}(t)-\phi x_{j}(t)
\end{align}
The variables $a_{ji}=\log(c\,w_{ji})$ quantify the reputation spillover from firm $j$ to firm $i$ via the ownership links, where $w_{ij}$ is the reported share firm $i$ holds in firm $j$ and $c$ is a normalization constant such that $a_{ji}$ is always equal or larger than one.
The second term in Eqn.~\eqref{eq:1} reflects the assumption that reputation fades out exponentially at a rate $\phi$ if it is not maintained. 

This dynamics has been also applied to model the reputation dynamics in online social networks \citep{Reptuation-OSN-2020}.
For application scenarios it is more convenient to use relative reputation values $r_{i}=x_{i}/x^{\mathrm{max}}$ instead of absolute values $x_{i}$.
But in this paper we are only interested in the reputation \emph{ranking} of firms, therefore we use $x_{i}$. 
We note that the dynamics of Eqn.~\eqref{eq:1} converges to an equilibrium value, quickly, which is then used for the ranking.

Once we have identified the set of driver nodes as described in the following section, we have to consider a control signal, i.e. a induced change that modifies the reputation of only the driver nodes.
The resulting linear dynamics can be conveniently expressed in matrix form: 
\begin{equation}
\dot{\mathbf X}(t)=\mathbf A^T \mathbf X(t) -\phi \mathbf X(t) + \mathbf B \mathbf U(t)
\label{eq1}
\end{equation}
The matrix $\mathbf{A}^{T}$ contains the information about the network topology,
the vector $X(t)=[x_1(t),x_2(t),...,x_N(t)]$ contains the reputation values of all firms. 
The vector $\mathbf{U}(t)  \in \mathbb{R}^{N_{c}}$ contains $N_c$ control signals $u_{k}(t)$ $(k=1,..,N_{c})$, and the matrix $\mathbf{B} \in \mathbb{R}^{N \times N_c}$ determines which firms are influenced directly by control signals.
That means, the elements $b_{ij} \neq 0$ if control signal $u_{j}(t)$ is applied to firm $i$.
To apply the concept of network controllability, we are still left with determining the set of driver nodes.

\subsection{Identification and classification of driver nodes}
\label{sec:ident-driv-nodes}

The recent framework  of structural controllability for complex networks \cite{Liu2011d} allows to identify minimum sets of driver nodes, i.e. a small number of nodes that can be utilized to control the whole network.
This method can be applied to directed networks.
Because we cannot repeat all details of the method here, we summarize the respective steps and refer to the literature for subsequent information \citep{Cornelius2013,Wang2012e,YZ-ACS,Peripherial-nodes-Zhang2016}.

A complex network of $N$ nodes can be controlled by different sets of driver nodes.
MDS denotes the \emph{minimum} set of drivers to control the \emph{whole} network, and the size of this set is $N_{d}$.
It is computationally infeasible to enumerate all the possible MDSs.
Therefore, in our paper we use two randomly chosen MDSs for the visualization in Figure~\ref{figure500} and calculate our control-related measures based on 10.000 random samples.

In different MDS we usually find different nodes, but some of them appear in every MDS.
The probability $P(D_{i})$ that a given node $i$ appears in an MDS is also known as control capacity $\mathcal{K}_{i}$  \citep{Jia2013f}.
Further, each driver $i$ controls a non-overlapping part of the network of size $N_{i}$.
The probability that a given node is in the subnetwork controlled by node $i$ is given by 
$P(N_{i})$.
We combine these two information in the conditional probability $P(N_{i}|D_{i})$ that a given node is part of the subnetwork controlled by $i$ given that $i$ is a driver node.
The upper bound of this probability is also known as \emph{control range}, $\mathcal{R}_{i}$ \citep{Wang2012e}.

To eventually combine control range and control capacity, we have proposed a new measure,  
\emph{control contribution} $C_{i}=\mathcal{K}_{i}\mathcal{R}_{i}$ \citep{YZ-ACS}.
This node-based measure gives us the the probability
for any node in a network to be controlled by node $i$ joint with the probability that $i$ becomes a driver.
Larger $\mathcal{C}_{i}$ indicate that  node $i$ is more important in driving the whole network to a desired state.
Concrete values for $\mathcal{C}_{i}$ can only be obtained algorithmically.
For an illustrative calculation and an algorithm we refer to Ref.~\citep{YZ-ACS}.
There, it was also demonstrated that control contribution is better suitable than control range or control capacity to classify the importance of nodes in controlling a network.

Applying the methods described above, we now have three different types of information for each firm in the ownership network: (i) its operating revenue $\Omega_{i}$, (ii) its reputation $x_{i}$ (which takes the weighted ownership relations $w_{ji}$ into account), and (iii) its control capacity $\mathcal{K}_{i}$, control range $\mathcal{R}_{i}$ and control contribution $\mathcal{C}_{i}$.
These measures reflect different dimensions to describe the importance of firms in an economic network, namely their economic activity, their dependence on other firms, and their influence on other firms.
Therefore, we can now address research questions that link these different dimensions, for instance: Are firms with a high operating revenue or firms with a high reputation also most influential in network control?

To quantify such relations, we perform an \emph{enrichment analysis}, a statistical method which is commonly used to identify genes or proteins that are over-represented \citep{Wuchty2014a}.
To illustrate the idea, suppose there are $N$ balls characterized by
colors $s$ and types $t$. 
We have three colors, i.e. $s$: (white, black, grey), and two types $t$: (heavy, light).
Enrichment analysis can, for example, tell whether heavy balls are more likely to be white balls or not.
To do so, we need to compare the number of heavy balls whose color is also white, $N^{s}_{l}$, and the number of heavy balls $N^{R}_{l}$, if we randomly sample $N/3$ balls. 

Here, we apply this analysis to the firms that are part of the driver set of size $N_{d}$. 
Our ``colors'', or categories, are now reputation values, i.e. $s:$ (low, medium, high) reputation.
To define these groups, we first calculate the reputation $x_{i}$ using Eqn.~\eqref{eq:1}, and then rank firms according to their reputation values in equilibrium.
The ranked set is split into three groups of equal size $N_{d}/3$.

Secondly, we specify which types $l$ we are interested in, for example
whether firms have a low, medium or high control contribution $C_{i}$. 
$N_l^{\mathrm{s}}\leq N_{d}/3$ then denotes the number of firms which are in the reputation group $s$ \emph{and} have a type $l$ regarding their control contribution. 
That means, instead of just looking into correlations across all firms, we define groups of firms with certain features and then address the question whether firms with these features appear more frequently than expected in each reputation group.

For this comparison, we need a random set R that has the same size $N_{d}/3$, but is sampled from all $N$ firms with respect to the feature $l$. 
$N^{\mathrm{R}}_{l}$ is the number of firms in the random set with, e.g., medium control contribution. 
The random sampling is performed 10.000  times, to obtain a distribution for the values $N_{l}^{R}$, from which we can calculate the mean $\mu(N_{l}^{R})$ and the standard deviation $\sigma(N_{l}^{R})$.
For the comparison between the category $s$ and the type $L$ we then use the $z$-score:
\begin{equation}
  \label{eq:2}
z_{l}^{s}=\frac{N^{s}_{l}\ \ -\mu(N_{l}^{R})}{\sigma(N_{l}^{R})}
\end{equation}
Obviously, a positive $z$-score shows an enrichment of the given category $s$ in the type $l$.
Enrichment means that firms with a given type $l$ appear more frequently in the category $s$ than expected at random.

\section{Results}
\label{sec:results}

\subsection{Driver nodes and access costs}
\label{sec:driv-nodes-1}

To classify firms as driver nodes, we first need to determine the size of the
MDS.
We find that, from the $1318$ firms in the ownership network, we need to control a minimum number of $N_d=341$ firms directly in order to control the whole network. 
Then, out of the large number of possible MDS with the same size, we have to
generate 10.000 random samples, on which our further analysis is based.

\begin{figure}[htbp] 
  \centering
\includegraphics[width=0.45\textwidth]{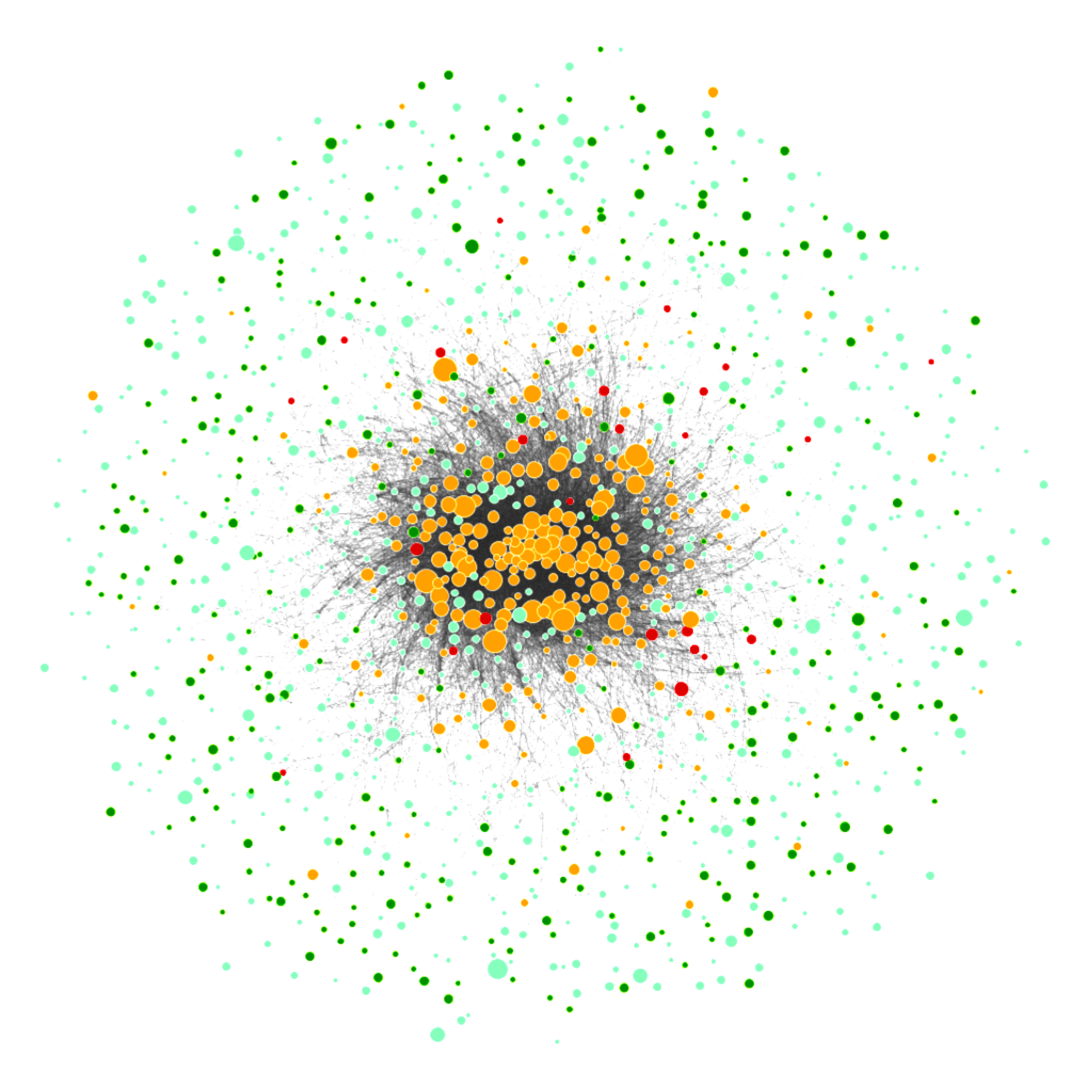}(a)
  \hfill
  \includegraphics[width=0.45\textwidth]{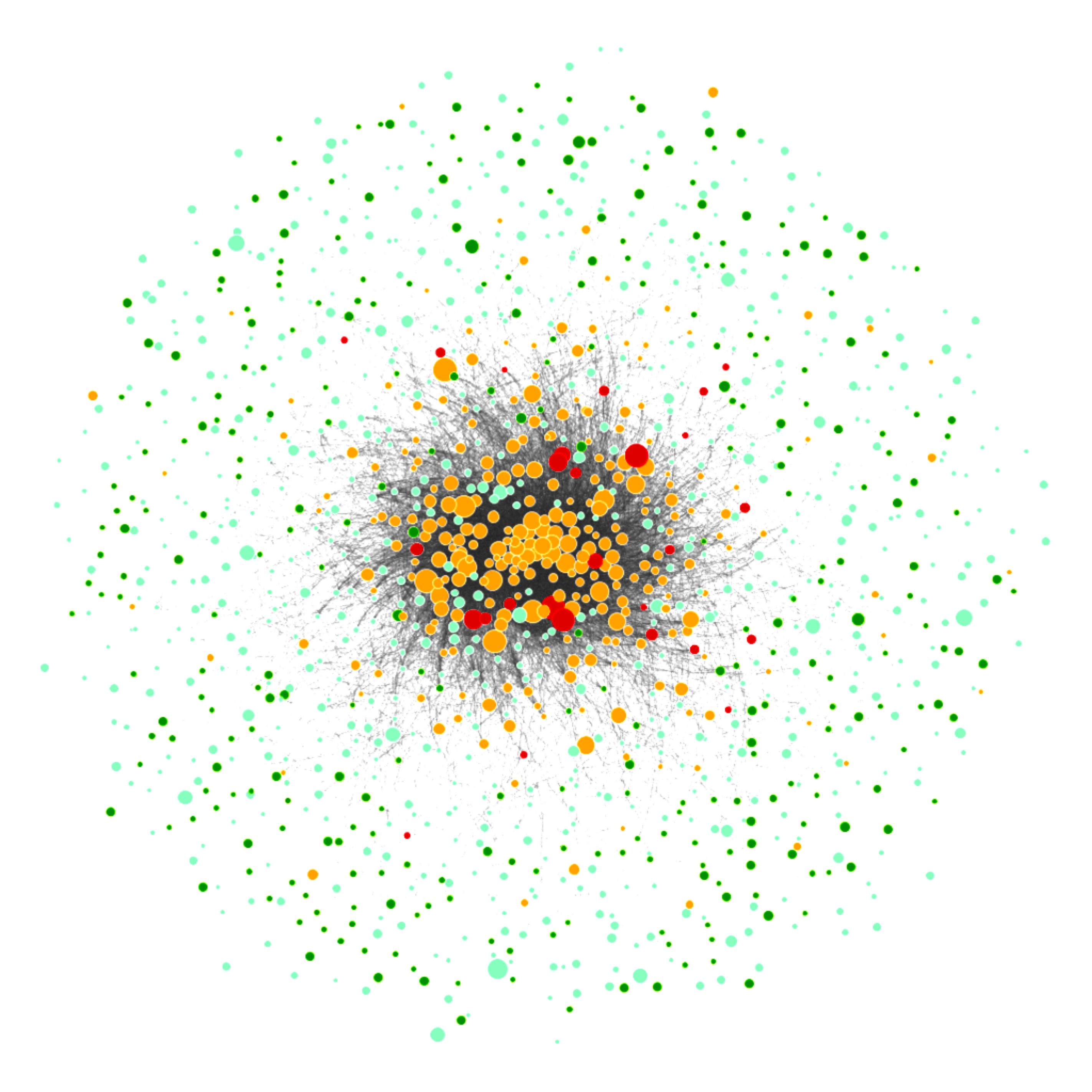}(b) 
  \caption{Visualizations of two MDSs in the ownership network. The node size is scaled proportional to the operating revenue of the firm. 
Transnational companies (TNCs) are shown in orange, and in red if they are also driver nodes. 
Participated companies (PCs) are shown in light green, and in dark green if they are also driver nodes.
  } 
  \label{figure500}
\end{figure}

As an illustration, Figure \ref{figure500} presents visualizations of two random MDSs embedded in the ownership network shown in Figure~\ref{fig:ownership}. 
We notice that both MDSs only have a few driver nodes in common.
Further, the right MDS contains more TNCs with high operating revenue as driver nodes, whereas the  left contains mostly PCs with lower operating revenue.

\begin{figure}[htbp] 
  \centering
\includegraphics[width=0.45\textwidth]{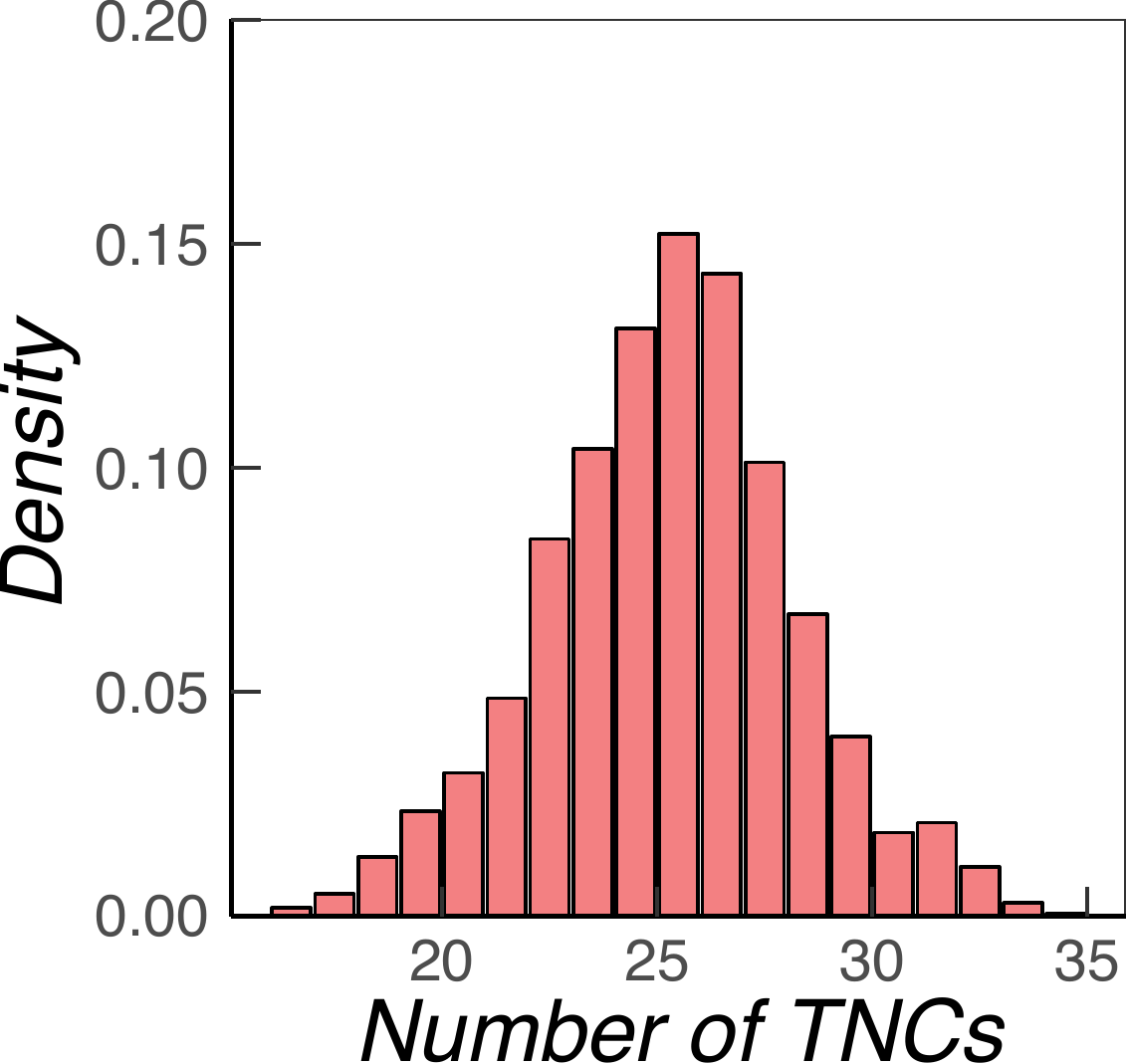}(a)
  \hfill
  \includegraphics[width=0.45\textwidth]{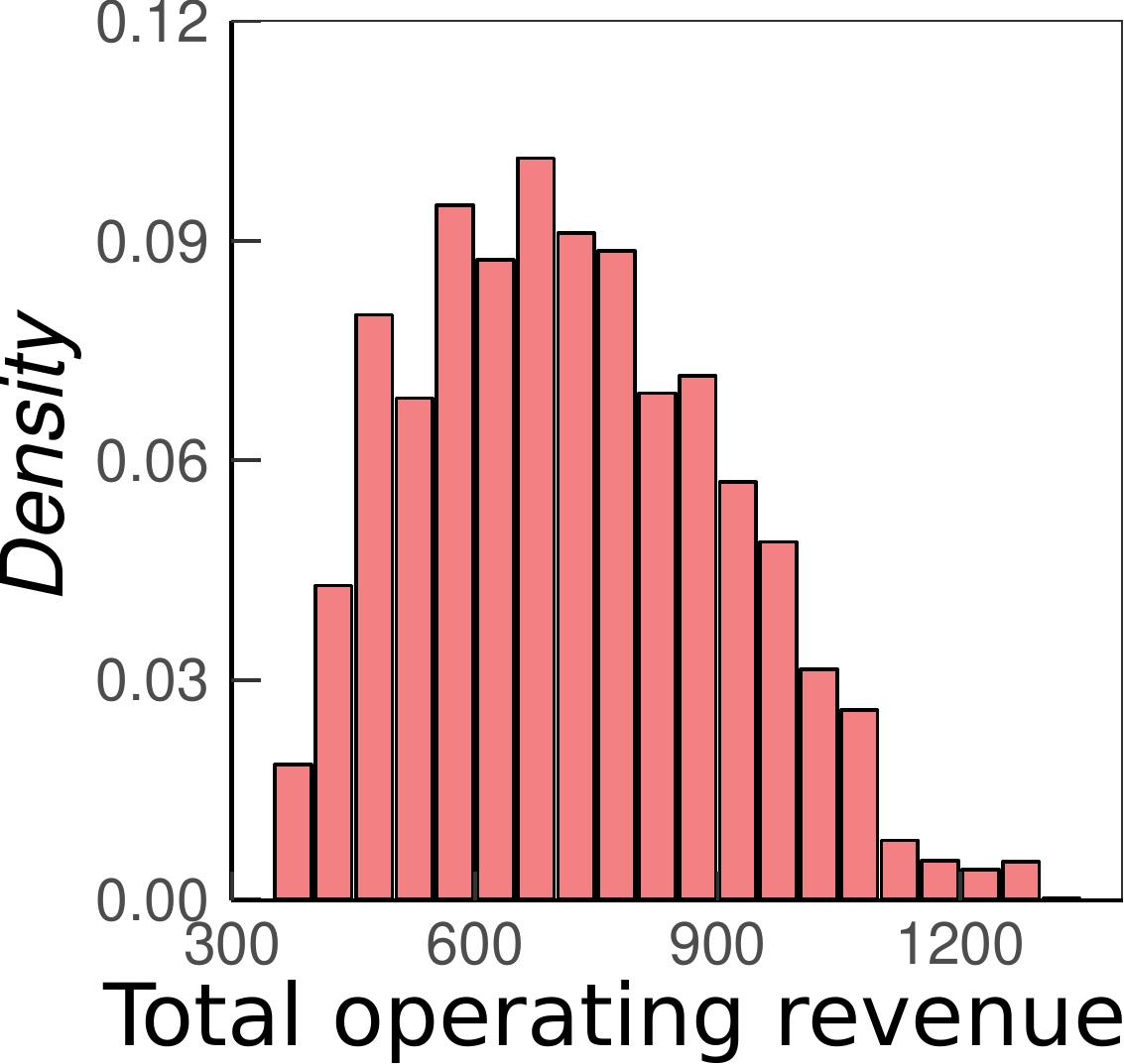}$\quad$(b) 
  \caption{Distribution of (b) the number of TNCs, and (b)
their total operating revenue (in trillion USD) sampled from 10.000 MDSs.   } 
  \label{figure501}
\end{figure}

To further quantify these differences, we first investigate how many TNCs are present in a randomly sampled MDS.
The distribution obtained from 10.000 MDS is shown in Figure~\ref{figure501}(b).
We find that on average about 26 TNCs are present in an MDS of size 341, i.e. less than 10\%.
One could na\"ively assume that because of their economic importance TNCs would be also the most important driver nodes and thus should appear more often. 
Interestingly, this is not the case.
Even more, the average of 26 TNCs, which corresponds to 8.7 \% of all TNCs in the core of the ownership network, is far below the \emph{expected} number of TNCs obtained from a \emph{random sample} of firms, which is 28.8\%.
This leads to the conclusion that TNCs are statistically underrepresented in the sets of driver nodes.

Further, the distibution is well formed  between a minimum of 17 and a maximum of 34 TNCs.
That means, we can find indeed MDSs in which the number of TNCs is only about 5\%. 
Why is this of interest?
These MDSs, because of the different number of TNCs, represent also a very different economic value, as proxied by the operating revenue $\Omega_{i}$ of their TNCs.
Figure~\ref{figure501}(a) shows the distribution of the $\sum \Omega_{i}$ of all TNCs in the 10.000 sampled MDSs. 
On average, the TNCs in an MDS hold a total operating revenue of 720 trillion USD, which accounts for 9.6\% of the amount held by all firms in the network.
But these values can be as low as 350 or as high as 1300 trillion USD.
So, we have a remarkable number of ``cheap'' MDS available. 

We remind that all MDS fulfill the same purpose, namely to control the \emph{whole} network.
But a ``cheap'' MDS, as proxied by the total operating revenue, with a low number of TNC would potentially be more easily accessible.
Remember that network controllability requires us to apply a control signal to a firm.
That means, we need to consider some sort of \emph{access cost} to utilize a given firm as a driver node.
It is likely more expensive to access a TNC of high operating revenue than a PC of low operating revenue.
Because we have no way to directly quantify the access cost, in the following we take the operating revenue $\Omega_{i}$  as a proxy of this access cost.

\begin{figure}[htbp] 
  \centering
  \includegraphics[width=0.5\columnwidth]{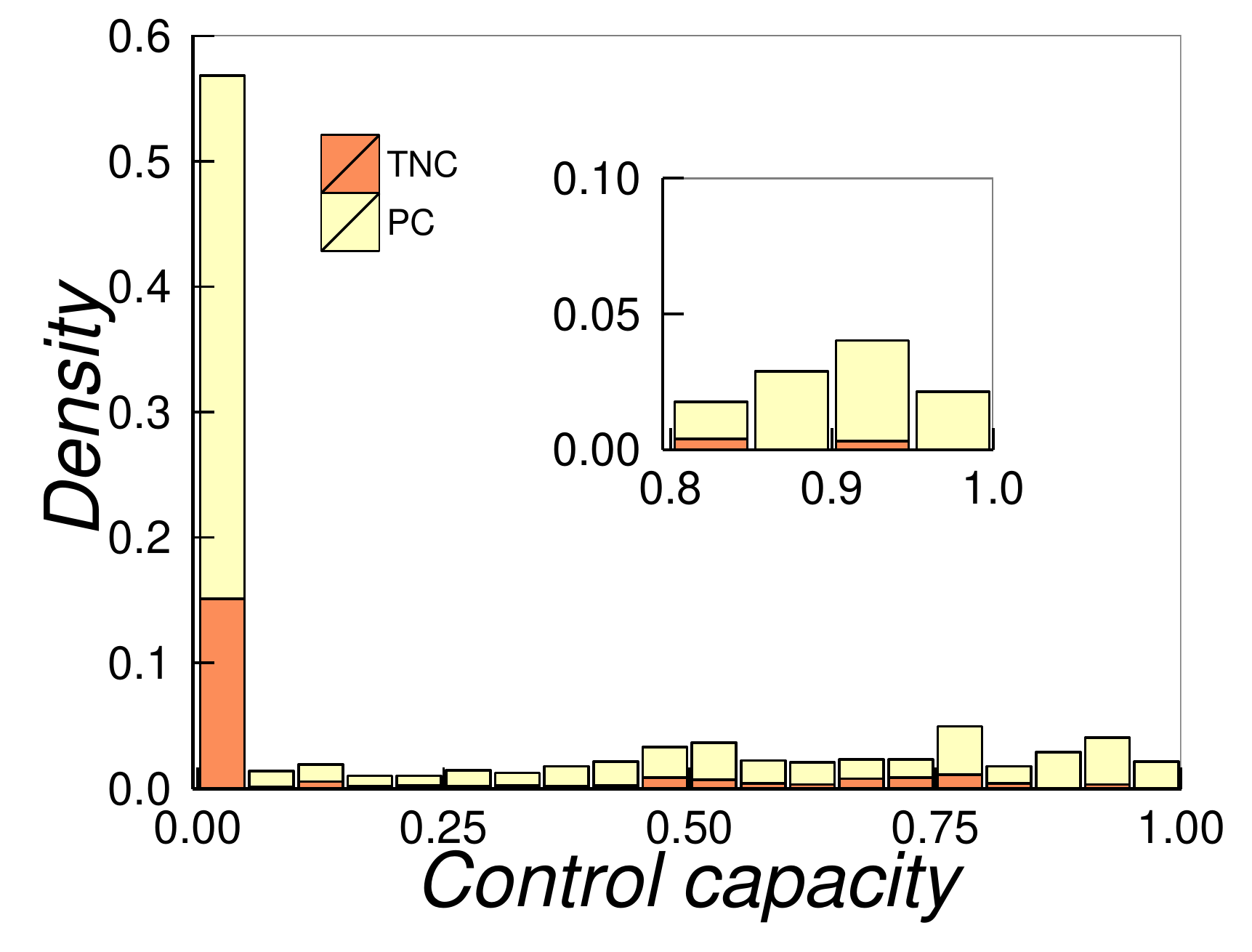} 
  \caption{Distribution of control capacities $\mathcal{K}_{i}$ for TNCs (red) and PCs (yellow). The inset enlarges the distribution around 1. } 
  \label{figure51}
\end{figure}

One could still argue that firms from a ``cheap'' MDS are less likely to be chosen as driver nodes, because they are more often PCs.
Again, this reflects the underlying assumption that TNCs should be more important as driver nodes and therefore should be also more often present in different MDSs. 
To refute this argument, we have investigated the distribution of the control capacities $\mathcal{K}_{i}$, which give the probability that a firm is chosen as a driver node. 
The results are shown in Figure \ref{figure51} both for TNCs and for PCs.
We find that most firms, despite they belong to a MDS, only have a very low probability to be chosen as driver nodes.
This holds for both TNCs and PCs.
Then, there is a very broad distribution of $\mathcal{K}_{i}$ values, which is largely dominated by PCs. 
Firms with a control capacity close to 1 are always present in any MDS.
We find that these firms are PCs.

Thus, in conclusion, firms that are PCs are most often present as driver nodes.
Secondly, their access cost should be considerably lower than for TNCs.
Therefore, we can safely choose ``cheap'' MDSs with a high fraction of PCs, to reach an efficient control of the whole network.
This is an important insight because it links network controllability to economic measures, and allows for policy advice.

\subsection{Different roles of nodes}
\label{sec:diff-roles-nodes}

So far, we have mainly explored the economic and control properties of the firms that are part of the sets of driver nodes.
Now we focus on the different \emph{types} $l$ of nodes, specifically the \emph{roles} of firms in (a) maintaining controllability, and (b) controlling the network. 
We start from our reputation ranking of firms, which lead to the formation of groups of size $N_{d}/3$ with $s:$ (low, medium, high) reputation, described in Section~\ref{sec:ident-driv-nodes}.

We first analyze how these groups correlate with the roles of firms in \emph{maintaining} control.
This requires us to specify the node types $l$ accordingly.
Maintaining control means that the set of driver nodes is still able to fully control the network, if a respective node $i$ would be removed.
Following \cite{Vinayagam2016}, we can then distinguish three types $l$ of driver nodes:
\setlist{nolistsep}
\begin{itemize}
\item[(a)] a node is \emph{indispensable} if after its removal \emph{more} driver nodes are needed; 
\item[(b)] a node is \emph{redundant} if its removal does not change the required number of driver nodes; 
\item[(c)] a node is \emph{dispensable} if after its removal the network is controllable with \emph{fewer} driver nodes. 
\end{itemize}
Based on this classification, to identify the role $l$ of firm $i$ in maintaining controllability we need to calculate the minimum number of driver nodes if  $i$ is removed, and compare it with the minimum number of driver nodes if $i$ is not removed.
We have to keep in mind that removing a node implies to change the topology of the network, which definitely impacts the size of the minimum set of drivers. An MDS of size 341 only holds for the non-perturbed network.

\begin{figure}[htbp] 
  \centering
\includegraphics[width=0.45\textwidth]{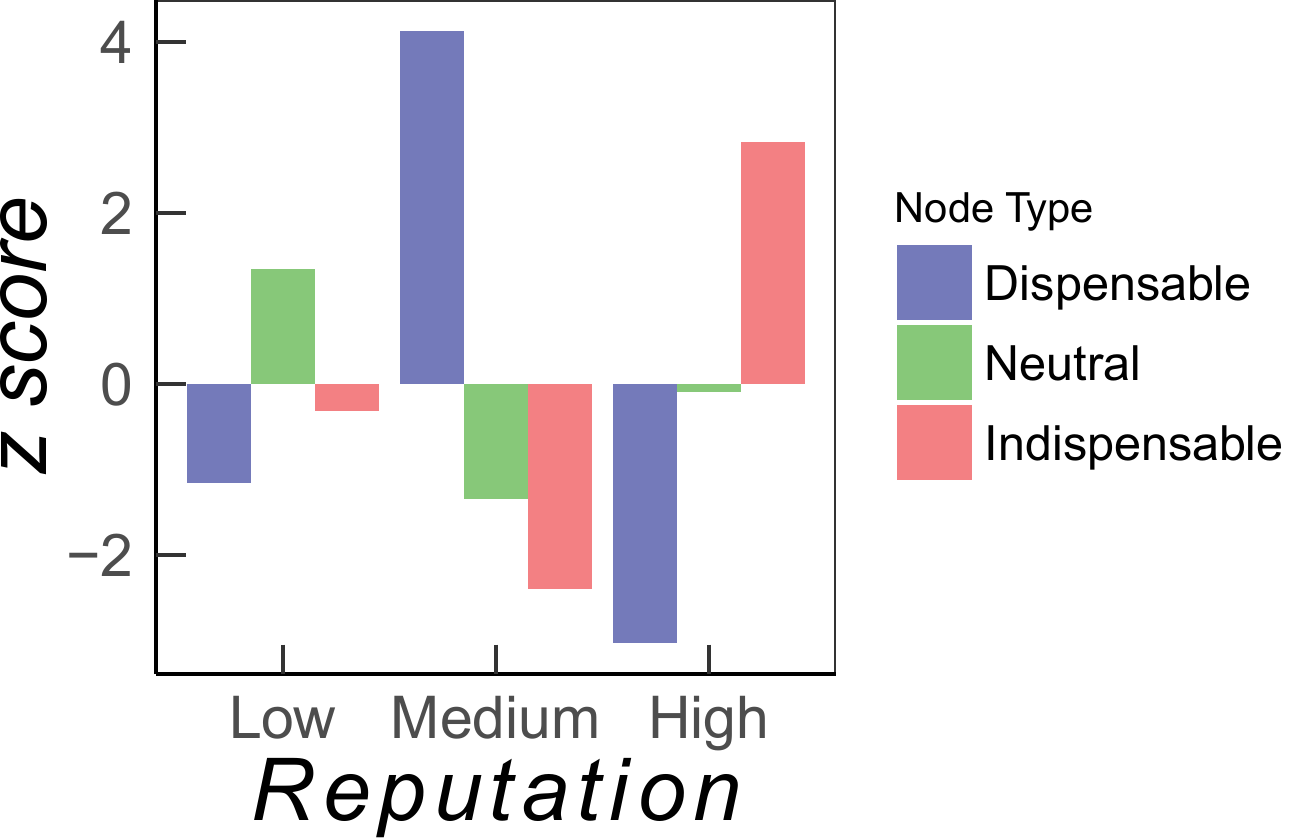}(a)
  \hfill
    \includegraphics[width=0.45\textwidth]{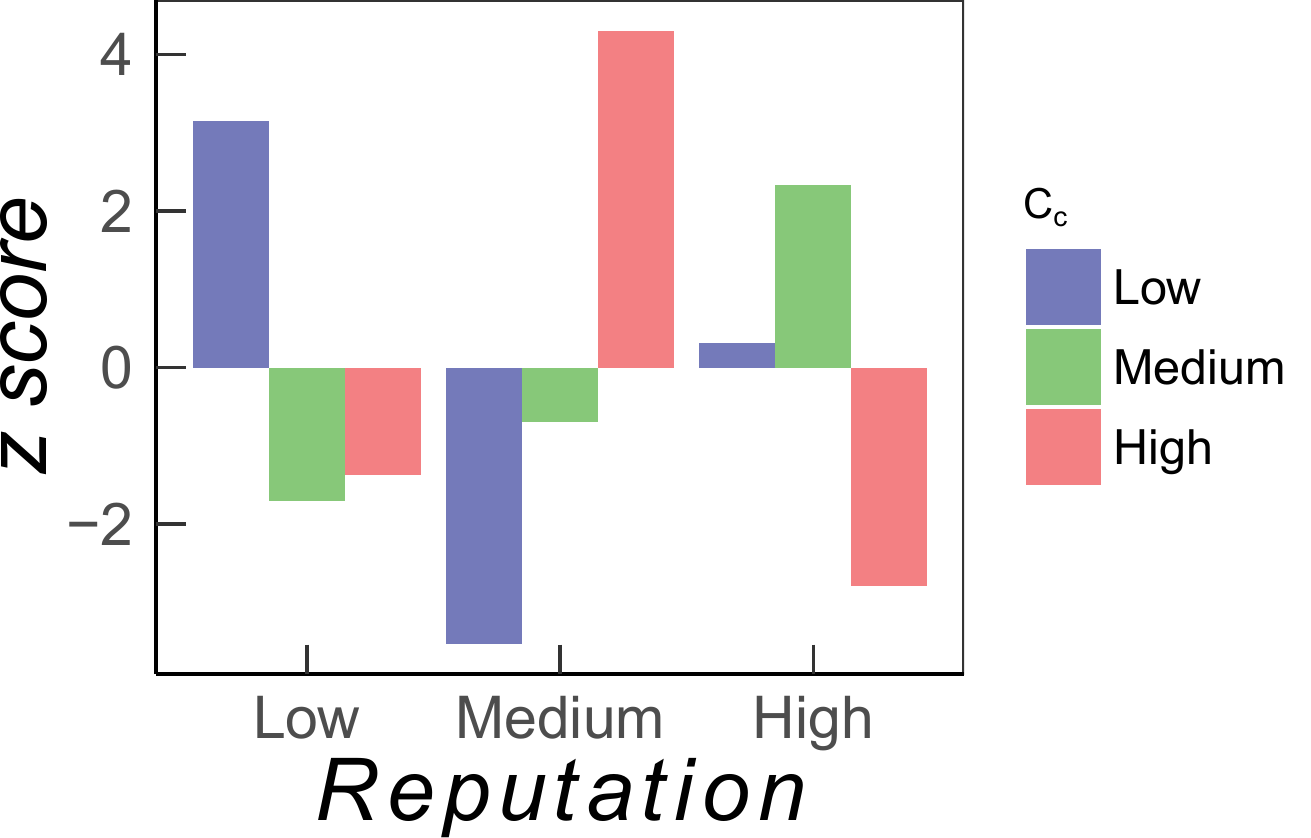}(b) 
  \caption{Enrichment analysis for firms classified according to their reputation value. (a) Role of firms in maintaining control, (b) role of firms with respect to control contribution, $\mathcal{C}_{i}$.} 
  \label{figure54}
\end{figure}

The results are shown in Figure \ref{figure54}(a) in terms of the $z$-score defined in Eqn.~\eqref{eq:2}.
Firms in each group with low, medium and high reputation do contain all three types of driver nodes, indispensible, redundant and dispensible.
But the $z$-score tells us whether such roles are enriched in a particular group.
We see that \emph{indispensable} nodes are most enriched in the set of high reputation firms.
This is in accordance with our expectation that the most reputable firms  channel control signals through the ownership network.
Interestingly, indispensable nodes are mostly underrepresented in the group with medium reputation instead of low reputation.
This can be partly explained from the fact that the ownership network forms  a strongly connected component.
Therefore, the removal of a low degree node, which is likely a firm of medium to low reputation, may leave some nodes with no incoming links, which have to be controlled directly with additional drivers.
In conclusion, this analysis shows  the importance of firms with high reputation in \emph{maintaining} controllability.

Secondly, we analyze how the three reputation groups correlate with the role of firms in \emph{controlling} the network.
In this case, we have to specify the types $l$ of nodes with respect to their  control contribution $\mathcal{C}_i$, introduced in Section~\ref{sec:ident-driv-nodes}.
We remind that $\mathcal{C}_i$ captures the probability for a firm to become a driver node, joint with the probability for any firm to be controlled by this firm.
Hence, firms with a high control contribution are top drivers.
We use the values of $\mathcal{C}_i$ to distinguish three groups of equal size $N_{d}/3$ with low, medium and high control contribution.

The results are shown in Figure \ref{figure54}(b) in terms of the $z$-score defined in Eqn.~\eqref{eq:2}.
We find that the top driver nodes are mostly firms with \emph{medium} reputation, not with high reputation, which is a very interesting result.
Firms with high reputation are strongly embedded into the ownership network and connected to other firms with high reputation.
Consequently, to utilize such firms as driver nodes would imply a considerable access cost.
But this is not needed.
Instead, network controllability can be best achieved with firms of medium reputation.

\begin{table}[]
\centering
\footnotesize
\begin{tabular}{clccr}
 \hline
$O_{\mathcal{C}}$ & Name                                                                                                              & Type & Country & $O_{x}$ \\
 \hline
1           & \begin{tabular}[c]{@{}l@{}}CAISSE REGIONALE DE CREDIT AGRICOLE \\ MUTUEL DE LA TOURAINE ET DU POITOU\end{tabular} & TNC  & FR                           & 592                   \\
\rowcolor[HTML]{EFEFEF} 
2           & BBVA CARTERA SICAV SA                                                                                             & PC    & ES                           & 1202                  \\
3           & BIOTECNET I MAS D SDAD ANONIMA                                                                                    & PC    & ES                           &1 198                  \\
\rowcolor[HTML]{EFEFEF} 
4           & INVERPASTOR SA SIMCAV.                                                                                            & PC    & ES                           & 1206                  \\
5           & INVERSIONES HERRERO SICAV SA                                                                                      & PC    & ES                           & 1201                  \\
\rowcolor[HTML]{EFEFEF} 
6           & BOLS HISPANIA SA SIMCAV.                                                                                          & PC    & ES                           & 1206                  \\
7           & \begin{tabular}[c]{@{}l@{}}BANQUE POPULAIRE LOIRE \\ ET LYONNAIS\end{tabular}                                     & PC    & FR                           & 551                   \\
\rowcolor[HTML]{EFEFEF} 
8           & \begin{tabular}[c]{@{}l@{}}CAISSE REGIONALE DE CREDIT AGRICOLE\\  MUTUEL DE NORMANDIE-SEINE\end{tabular}          & PC    & FR                           & 578                   \\
9           & \begin{tabular}[c]{@{}l@{}}BANQUE POPULAIRE BOURGOGNE\\  FRANCHE-COMTE\end{tabular}                               & PC    & FR                           & 530                   \\
\rowcolor[HTML]{EFEFEF} 
10          & BANQUE POPULAIRE DU NORD                                                                                          & PC    & FR                           & 542                   \\
            &                                                                                                                   &      

\end{tabular}
\caption{
  List of the firms that are the top 10 driver nodes ranked by their control contributions, $\mathcal{C}_{i}$. $O_{\mathcal{C}}$ denotes the respective rank.
  $O_{x}$ denotes the rank of the same firm with respect to the reputation $x$.
}
\label{table31}
\end{table}

In Table \ref{table31} we also list the top 10 driver nodes with respect to their control contribution $\mathcal{C}_{i}$ and provide their reputation rank. 
We observe that none of these firms has a high reputation, and only one of them is a TNC.
This confirms that the top drivers are likely not firms with high reputation in the ownership network.

To summarize our findings from the two enrichment analyses: (a) Firms with  high reputation maintain the controllability of the network, but are unlikely to become top driver nodes. (b) Firms with medium reputation are most likely to become top driver nodes, but they are also
dispensible for maintaining controllability.

\section{Discussion}
\label{sec:discussion}

Our analysis makes two major contributions to the state-of-the-art in network science:
(i) we provide new ways of quantifying the importance of firms, and
(ii) we link two strands of research that are so far largely disconnected: network controllability and economic networks.
In the following, we comment on these achievements.

Starting from network science, the importance of nodes in a network is most often quantified based on the network topology.
This is reflected in different centrality measures \citep{das2018study,landherr2010critical}, which have been recently extended also to temporal networks \citep{Scholtes2016}.
Such a characterization, by design, neglects the fact that networks serve a purpose, links have a meaning, nodes have a intrinsic dynamics.

To cope with this, we need importance measures that consider the \emph{dynamics on the network}.
There is no general ``importance'', but importance with respect to a given process that we want to describe.
Our application scenario is \emph{reputation spillover}.
This requires us to quantify (a) the reputation of firms, and (b) the process of reputation spillover.
For this, we have utilized a recent framework to model reputation dynamics \citep{Reptuation-OSN-2020}.
But, to become relevant and applicable, this approach needs an \emph{economic} interpretation.
This problem was also solved in a recent study that links reputation spillovers to ownership relations \citep{Zhang2019}.
That means, at this point we have a new way to quantify the importance of firms by means of a reputation value that reflects ownership relations.
This complements other importance measures for firms, such as their operating revenue.

In this study, we go one step further, by linking these importance measures to the role of firms in network control.
Using the \emph{topology} of the ownership network and the \emph{dynamics} of reputation spillover, we can apply the recent concept of network controllability \citep{Liu2011d,Cornelius2013}. 
It allows to identify those firms that can become \emph{driver nodes} to steer the reputation dynamics.
We find that, out of the 1318 firms that form the core of the ownership network, an MDS of only 314 firms, i.e. 23.8\%, less than one quarter, is needed to control the dynamics of the \emph{whole} network.
To characterize the control contribution of each firm, we have calculated a new measure $\mathcal{C}_{i}$ \citep{YZ-ACS}.
It combines two information, the probability of a firm to become a driver and the probability that other firms are controlled by firm $i$.

Hence, we now have two importance measures, in addition to the operating revenue $\Omega_{i}$, the reputation value $x_{i}$ and the control contribution $\mathcal{C}_{i}$.
Each of these measures reflects a different dimension: economic activity, dependence on other firms and influence on other firms.
This eventually enables us to better characterize those firms that are most important in controlling the reputation dynamics.

Precisely, our enrichment analysis tells whether firms of low, medium or high reputation are more often than expected involved in maintaining or exerting control.
Again, one could na\"ively expect that large firms with a high operating revenue, such as TNCs, or firms with the highest reputation play the most important role.
As our analysis shows, this is not the case.
TNCs are underrepresented in the minimum sets of driver nodes, which are dominated by PCs.
And firms with a high reputation are less likely to become top driver nodes.
Instead, we find that firms with medium reputation play the most important role as top drivers.

This is not an abstract insight, it can be given an economic interpretation, this way linking network controllability and economic networks.
The nodes of our network are not abstract entities, they are economic actors characterized by their ownership structure, $\omega_{ij}$ and their operating revenue, $\Omega_{i}$.
This enables us to distinguish between transnational companies (TNCs) and participated companies (PCs).
This information can be used to argue about \emph{access costs}, i.e. the potential costs if one wants to use specific firms as driver nodes.

Network controllability implies that control signals need to be applied to certain nodes.
Hence, in an economic setting there are costs involved, not only for the control signal, but also for accessing the node.
As we demonstrate, among the various sets of driver nodes that control the whole network, there are many MDSs comprised of PCs of a lower total operating revenue.
If operating revenue is taken as a proxy for the access cost, these MDS would be quite ``cheap'' to access, while still allowing for full control.
A similar argument holds for firms with high reputation, which are likely TNCs with a high operating revenue.
As we have shown, firms of medium reputation play the major role in controlling the network.
These are mostly PCs with a lower operating revenue and, hence, with a lower access cost.

In conclusion, using these economic criteria we can select sets of driver nodes that are less costly to access, but still allow for a full control of the network.
This is an interesting finding because it opens new ways of discussing the  \emph{economic importance} of firms.

\small \setlength{\bibsep}{1pt}

\end{document}